\documentclass[%
reprint,
showpacs,preprintnumbers,
bibnotes,
 amsmath,amssymb,
 aps,
]{revtex4-1}

\usepackage{graphicx}
\usepackage{dcolumn}
\usepackage{bm}
\usepackage{url}

\newcommand\aj{Astron. J.}

\newcommand\aap{Astron. Astrophys.}

\newcommand\app{Astropart. Phys.}

\newcommand\apjl{Astrophys. J. Lett.}

\newcommand\cqg{Class. Quantum Gravity}

\newcommand\jcap{J. Cosmol. Astropart. Phys.}

\newcommand\lrr{Living Rev. Relativ.}

\newcommand\mnras{Mon. Not. R. Astron. Soc.}

\begin{document}

\title{Astrometric test of the weak equivalence principle}

\author{Yi Xie}
 \email{yixie@nju.edu.cn}
\affiliation{
School of Astronomy and Space Science, Nanjing University, Nanjing 210093, China\\
Key Laboratory of Modern Astronomy and Astrophysics, Nanjing University, Ministry of Education, Nanjing 210093, China
}

\date{\today}

\begin{abstract}
	Weak equivalence principle (WEP) is, for the first time, tested by astrometry on quasars in the sky measured in two wavelengths. Compared to previous WEP tests based on the Shapiro time delay of massless particles, this one has profound superiority that nearly \mbox{1 700} quasars with best measured positions commonly in the optical and radio bands are available. It ensures that, among the tests with photons, this one can give the most significantly robust bound on possible violation of WEP.   
\end{abstract}


\maketitle

\allowdisplaybreaks

\textit{Introduction.} -- Belonging to ternary pieces of the Einstein equivalence principle, the weak equivalence principle (WEP) states that the trajectory of a freely falling test body does not depend on its composition and internal structure \cite{Will1993TEGP,Will2014LRR17.4}. It was recently tested by a space-borne E\"otv\"os experiment with two test masses \cite{Touboul2017PRL119.231101}. It can also be tested by massless particles with different properties such as unequal frequencies in the context of the parameterized post-Newtonian (PPN) formalism and any possible violation of WEP is characterized by the difference between values of the PPN parameter $\gamma$ of these massless particles \cite{Will1993TEGP}. Such a frequency-dependence of $\gamma$ can arise in quantum gravity but is also expected to be suppressed by powers of the ratio of the frequency to the Planck mass, making it hardly observable \cite{Amelino-Camelia2013LRR16.5}. Nevertheless, some theories \cite[e.g.][]{Ellis2004AstPar20.669,Accioly2001PRD64.067701,Magueijo2004CQG21.1725} argue that it might manifest at the scale that is many orders of magnitude lower than the Planck scale.

Pioneered by supernova 1987A tests of WEP \cite{Longo1988PRL60.173,Krauss1988PRL60.176}, the $\gamma$-dependent Shapiro time delay has been intensively employed nowadays \cite[e.g.][]{Wei2015PRL115.261101,Wang2016PRL116.151101}. However, the number of available astroparticle and high energy events for this particular kind of test is very limited, maximally up to $\mathcal{O}(20)$ \cite{Sang2016MNRAS460.2282}. Therefore, unlike the E\"otv\"os experiment \cite{Touboul2017PRL119.231101}, the outcomes of Shapiro delay tests are lack of significantly statistical robustness even though some upper limits of the violation were found to be extremely small \cite{Yang2016PRD94.101501,Wu2017PRD95.103004}. 

In this work, a new test of WEP by astrometry is presented. Rather than focusing on Shapiro time delay, I consider directions of incoming photons. This direction contains unperturbed part and relativistic light bending which depends on $\gamma$ in the PPN formalism \cite{Will1993TEGP}. Any difference between the directions of one source respectively measured in different wavelengths might indicate possible violation of WEP which can be described by differential value of $\gamma$.  

To fulfill such an astrometric test, two officially published quasars catalogues with the highest precision and accuracy in the optical and radio bands are used. Recently, \textit{Gaia} Data Release 2 (GDR2) was just available \cite{GaiaCol2018A&A.1}. The celestial reference frame of GDR2 (GCRF2) is defined by more than half a million extragalactic sources with unprecedentedly low uncertainties in their positions, making it the best reference frame in the optical band today \cite{Mignard2018}. More than two thousand sources in GCRF2 have radio counterparts in the second realization of the International Celestial Reference Frame (ICRF2) \cite{Ma2009ITN35.1,Fey2015AJ150.58} which is the best openly available reference frame in the radio band now. The optical positions match to the radio ones at the level well below sub-milliarcsecond for most common sources \cite{Mignard2018}. The number of these common sources with best positions overwhelms the size of datasets of Shapiro time delay tests and guarantees much more significantly statistical robustness for the astrometric test.

\textit{Theory.} -- For a static observer at the barycenter of the Solar System, a given extragalactic source in one of the catalogues with coordinates right ascension and declination $(\alpha, \delta)$ has its observed unit coordinate direction as
\begin{eqnarray}
  \label{}
  -\bm{n}  =  ( \cos \delta \cos \alpha , \cos \delta \sin \alpha, \sin \delta )^{\mathrm{T}},
\end{eqnarray}
where the aberration due to the observer's motion and the light deflection caused by the gravitational bodies in the Solar System (that is dominated by the Sun) have been properly corrected and removed in the processing of observational data based on standard relativistic astrometric models for constructing the catalogue \cite[e.g.][]{Klioner2003AJ125.1580}. However, light deflections caused by gravitational bodies outside the Solar System are not taken into account. In the PPN formalism, the (negative) unit direction has two contributions:
\begin{equation}
  \label{}
  \bm{n}=\bm{\sigma}+\delta\bm{\sigma}_{\mathrm{pN}},
\end{equation}
where $\bm{\sigma}$ is the unperturbed unit tangent vector of the light ray at the emission and $\delta\bm{\sigma}_{\mathrm{pN}}$ is the sum of all post-Newtonian gravitational effects due to the bodies outside the Solar System on the photon's trajectory. It makes a big difference between astrometric and time delay tests because light deflection is not a cumulative effect. Since different gravitational bodies can deflect the photon in various directions, the deflection effect will not necessarily grow for far away sources, which is contrary to time delay. Therefore, such a circumstance demands very precise modeling of the gravitational potential on the light of sight all the way from the source to the observer. The noncumulative property of light deflection seems to make astrometry not very much sensitive for the test; however, plenty of all-sky distributed sources with increasingly improved positions might alleviate this restriction. In practice, it would be extremely difficult to include all of gravitational potentials into the theoretical consideration which might introduce a lot of uncertainties. While the inclusion of an \textit{averaged} gravitational potential fluctuation from the large-scale structure can improve the results from time delay tests \cite{Nusser2016ApJ821.L2}, the noncumulative property renders the resulting improvement from such an inclusion for an astrometric test undetermined that will be left for future detailed investigation. Therefore, only most massive and significant gravitational deflectors will be taken into account here. 

If it is assumed that the largest contribution to $\delta\bm{\sigma}_{\mathrm{pN}}$ come from the spherically symmetric components of the gravitational fields of these most massive bodies, it can be obtained as (in geometrized units of $G=c=1$) \cite{Klioner2003AJ125.1580}
\begin{equation}
  \label{}
  \delta\bm{\sigma}_{\mathrm{pN}} = -(1+\gamma) \sum_{\mathrm{A}} M_{\mathrm{A}}\frac{\bm{d}_{\mathrm{A}}}{|\bm{d}_{\mathrm{A}}|^2} \bigg(1 + \bm{\sigma} \cdot \frac{\bm{r}_{\mathrm{oA}}}{|\bm{r}_{\mathrm{oA}}|} \bigg),
\end{equation}
where $M_{\mathrm{A}}$ is the mass of body A; $\bm{r}_{\mathrm{oA}}=\bm{x}_{\mathrm{o}}-\bm{x}_{\mathrm{A}}$, and $\bm{x}_{\mathrm{o}}$ and $\bm{x}_{\mathrm{A}}$ are respectively the positions of the observer and the body A; and $\bm{d}_{\mathrm{A}}=\bm{\sigma} \times (\bm{r}_{\mathrm{oA}} \times \bm{\sigma})$. As the closest and most massive gravitational bodies, the Milky Way, the Virgo Cluster and the Laniakea Supercluster of galaxies \cite{Tully2014Nature513.71} are adopted as the deflectors and denoted respectively with subscripts ``M'', ``V'' and ``L'' for short. They were also chosen as the gravitational bodies in the Shapiro delay tests \cite{Longo1988PRL60.173,Krauss1988PRL60.176,Wei2015PRL115.261101,Wang2016PRL116.151101,Wei2016JCAP08.031,Wu2017PRD95.103004}. The mass, distance and coordinates of the Milky Way are taken as $M_{\mathrm{M}}=6\times10^{11}$ $M_{\odot}$ \cite{Kafle2012ApJ761.98}, $r_{\mathrm{oM}}=8.32$ kpc \cite{Gillessen2017ApJ837.30} and $(\alpha_{\mathrm{M}},\delta_{\mathrm{M}})=(17^{\mathrm{h}}45^{\mathrm{m}}40.^{\mathrm{s}}0409,-29^{\circ}00'28."118)$ \cite{Reid2004ApJ616.872}. While the Virgo Cluster with $M_{\mathrm{V}}=1.2\times10^{15}$ $M_{\odot}$ is a part of the Laniakea Supercluster with total mass $M_{\mathrm{L}}=10^{17}$ $M_{\odot}$, its direction $(\alpha_{\mathrm{V}},\delta_{\mathrm{V}})=(12^{\mathrm{h}}27^{\mathrm{m}},12^{\circ}43')$ and distance $r_{\mathrm{oV}}=16.5$ Mpc are both far away from the center of Laniakea $(\alpha_{\mathrm{L}},\delta_{\mathrm{L}})=(10^{\mathrm{h}}32^{\mathrm{m}},-46^{\circ}00')$ and $r_{\mathrm{oL}}=77$ Mpc so that they are separately treated. This point-mass approximation of these gravitational potentials was assessed to be valid for sources far from the deflectors \cite{Longo1988PRL60.173,Krauss1988PRL60.176,Wei2016JCAP08.031,Wu2017PRD95.103004}. 

Intergalactic and galactic media can also cause wavelength-dependent light bending. Although the detailed all-sky map of refractive indices is barely known, it is reasonably expected that the dispersive effects of the Galactic medium decreases roughly with the increment of distance to the Galactic plane. Before a well-established dispersion map becomes available, its noise has to be mitigated by removal of sources very close to the plane.

With measurements in \textit{Gaia}'s unfiltered optical $G$ band and in the ICRF2's radio band, the unit direction $\bm{n}$ has its respective values $\bm{n}(G)$ and $\bm{n}(\mathrm{R})$ whose difference yields $\gamma(G)-\gamma(\mathrm{R})$, indicating any possible violation of WEP.

\textit{Datasets.} -- The ESA \textit{Gaia} mission is mapping the sky in the optical band \cite{GaiaCol2016A&A595.A1}. Based on observations collected during the first 22 months of operational phase, its second data release GDR2 provides astrometry for more than 1.3 billion sources and radial velocities for more than 7.2 million stars \cite{GaiaCol2018A&A.1}. On the contrary to its first data release (GRD1) \cite{GaiaCol2016A&A595.A1,GaiaCol2016A&A595.A2}, such an astrometric solution no longer depends on the Tycho-2 Catalogue. It is also improved with median positional uncertainties of 0.04 milliarcsecond (mas) for bright sources and of 0.7 mas for faint sources \cite{Lindegren2018}. Its celestial reference frame, GCRF2, consists of the positions of \mbox{556 869} sources in GDR2 with median positional uncertainties of 0.12 mas for $G < 18$ mag and of 0.5 mas at $G = 20$ mag \cite{Mignard2018}. Belonging to these sources, \mbox{2 327} ones match to a subset of ICRF2 sources in the radio band. ICRF2 contains \mbox{3 414} radio quasars observed by very long baseline interferometry with an accuracy floor of 40 microarcsecond ($\mu$as) \cite{Ma2009ITN35.1,Fey2015AJ150.58}. These two frames respectively realized in the optical and radio bands globally agree with each other at the level of several tens of $\mu$as \cite{Mignard2018}.  For the majority of the common sources, their optical positions match to the radio ones at the level lower than 1 mas or even better; however, there exist some discrepant sources \cite{Mignard2018}.

These optical-radio offsets arise for a number of intrinsic and extrinsic reasons even without any violation of WEP. In fact, it could be expected that such ``noise'' is much louder. The offsets between centroids of radio and optical emission can have their astrophysical origins due to jet structure of quasars and synchrotron opacity \cite{Kovalev2017A&A598.L1,Petrov2017MNRAS471.3775} in which the underlying physics is still little known. It was found \cite{Mignard2016A&A595.A5,Makarov2017ApJ835.L30,Petrov2017MNRAS467.L71} that double sources, confusion sources, pronounced extended structures, dust structures and surrounding bright distribution of light can also account for these offsets. A case study on about 10 discrepant sources commonly in GDR1 and ICRF2 confirmed that their positions are damaged by poor observation due to low brightness and false detection due to confusion with nearby brighter sources \cite{Mignard2016A&A595.A5}, but origins of nearly 200 other discrepant sources remain unclear. For the discrepant sources commonly in GCRF2 and ICRF2, a further investigation on their individual causes is still unavailable. It would be very challenging to completely eliminate this contamination since these astrophysical and environmental disturbances can hardly be well-handled at least in the current stage. Therefore, the whole sample of \mbox{2 327} common sources between GCRF2 and ICRF2 is adopted for the astrometric test in order to prevent an artificial and biased outcome towards a null-result from prior choosing sources with good agreement in the optical and radio bands. The systematics uncertainties caused by these disturbances will be assessed and estimated by a nonparametric statistical method in the next section.   

This sample needs to be further narrowed down in order to ensure the point-mass approximation for the deflectors and to reduce the dispersive effects. Hence, sources are chosen according to their angular distances to the deflectors beyond 15 degrees based on the angular sizes of the deflectors and according to their Galactic latitude larger than 20 degrees for avoiding the Galactic plane. Such limits seem to be somewhat conservative. It is no doubt that after fine-tuning the sample of sources and criteria for choosing sources, a better bound on the violation of WEP can be obtained, which is, however, out of the scope of this work. 

After filtering with the limits of point-mass approximation for the deflectors and of the sources' Galactic latitude, there are finally selected \mbox{1 697} common sources in GCRF2 and ICRF2 left for statistical inference on the violation of WEP.

\textit{Results.} -- It is straightforward to estimate $\gamma(G)-\gamma(\mathrm{R})$ and its standard deviation based on the aforementioned dataset by the weighted least-squares method. However, such an estimator and its statistical uncertainty might be biased due to the absence of systematic uncertainties in the statistical inference. It is important to assess the systematics which originate from theoretical and observational aspects.   

Systematics from theoretical unmodeling and mismodeling and the ways to mitigate them in the data preparation have been addressed previously, while systematics from observations are much more complicated. It was claimed \cite{GaiaCol2016A&A595.A1} that it has to wait towards the end of the \textit{Gaia} mission when all calibration will be successfully handled so that its systematic effects can be controlled down to $\mu$as level. Meanwhile, independent analyses on ICRF2 with different softwares found that some sources in the southern hemisphere suffer from ``declination bias'' up to a few hundreds of $\mu$as, for which the reason is still not fully understood (see \cite{Mayer2017A&A606.A143} for a recent discussion).

Therefore, nonparametric statistical methods \cite{Wasserman2006} for computing standard errors and confidence intervals could be an appropriate way to deal with these poorly-known systematics until knowledge of them is dramatically improved. The jackknife is a resampling method for estimating the bias and variance of an estimator. The bias of the estimator can be estimated from differences between the estimator with all of observations and the ones with some observations removed \cite{Quenouille1949JRSSB11.18,Tukey1958AMS29.614}. The jackknife method was recently employed in the lunar laser ranging experiments on the standard-model extension for estimating the contributions of systematics \cite{Bourgoin2016PRL117.241301,Bourgoin2017PRL119.201102}.

More specifically, the delete-$m_j$ jackknife method \cite{Busing1999StatCom9.3} is taken. This method is valid when the sample is divided into $g$ groups with different sizes. It is supposed that $\hat{\theta}_n$ is an estimator of a parameter $\theta$ based on the sample of $n$ observations. The bias-corrected estimators of $\theta$ and its variance are   
\begin{equation}
  \label{}
  \hat{\theta}_{J(m_j)} =  g \hat{\theta}_n-\sum^g_{j=1}\bigg(1-\frac{m_j}{n}\bigg)\hat{\theta}_{(j^*)}
\end{equation}
and
\begin{eqnarray}
  \label{}
  \hat{\sigma}^2_{J(m_j)} & = & \frac{1}{g}\sum^g_{j=1}\frac{1}{h_j-1}\bigg[h_j\hat{\theta}_n-(h_j-1)\hat{\theta}_{(j^*)}\nonumber\\
  & & -g\hat{\theta}_n + \sum^g_{k=1}\bigg(1-\frac{m_k}{n}\bigg)\hat{\theta}_{(k^*)}\bigg]^2
\end{eqnarray}
where $\hat{\theta}_{(j^*)}$ is an estimator of $\theta$ based on a sample with $m_j$ observations in the group $j$ removed and $h_j=n/m_j$. 

In the present analysis, the estimator is the weighted least-squares estimator, the parameter is $\gamma(G)-\gamma(\mathrm{R})$ and the whole sample is all of the finally selected common sources in GCRF2 and ICRF2 with $n=$\mbox{1 697}. The all sky is divided into 8 equal areas: four in the southern hemisphere and four in the northern one. In each hemisphere, each area covers 6 hours in the right ascension from $0^{\mathrm{h}}$ to $24^{\mathrm{h}}$. According to the area which a source belongs to, the dataset can be divided into 8 groups for the delete-$m_j$ jackknife.

For the finally selected common sources in GCRF2 and ICRF2, it is obtained that
\begin{equation}
  \label{}
  \gamma(G)-\gamma(\mathrm{R})= (-5.8\pm0.3|_{\mathrm{stat.}}\pm4.4|_{\mathrm{sys.}})\times10^{-6},
\end{equation}
where the estimator is given by the weighted least-squares method and its statistical and systematic uncertainties are respectively estimated by the weighted least-squares method and the jackknife method. The estimator by the jackknife method based on the same sample is $\gamma(G)-\gamma(\mathrm{R})= -6.8\times10^{-6}$. While the estimators from these two methods are consistent, its systematics uncertainty is over 10 times larger than the statistical one, implying the existence of possible incompleteness in the theoretical model and uncleaned interference in the dataset. 

As a preliminary demonstration of effect due to the optical-radio offsets probably caused by barely known systematics, a subset with \mbox{1 493} sources is chosen from the finally selected common sources by the unnormalized and normalized optical-radio angular differences within 10 mas and 4.1 in which the normalized difference is dimensionless (see \cite{Mignard2016A&A595.A5} for details). With the same statistical methods, this subset yields $\gamma(G)-\gamma(\mathrm{R})= (-3.2\pm0.3|_{\mathrm{stat.}}\pm0.5|_{\mathrm{sys.}})\times10^{-6}$ where the jackknife method gives almost the same estimator as the weighted least-squares one. Despite the decrement of its size by loss of about 200 sources, it gives a smaller estimator and much less systematics in the statistical sense. Although the outcome from such a refined subset is biased and only considered as a consistency check of the bound obtained before, it suggests that sources with large optical-radio offsets require further detailed case studies one by one.

One interesting and intermediate result is that, during the jackknifing the finally selected common sources, one estimator of $\gamma(G)-\gamma(\mathrm{R})$ with sources in the southern hemisphere removed is smaller by a factor of $2$ than the ones with sources in the northern excluded, which indirectly supports the existence of ``declination bias''.

\textit{Conclusions.} -- WEP is, for the first time, astrometrically tested by extragalactic sources in the sky measured in the two wavelengths. The profound superiority of such a test is that nearly \mbox{1 700} sources with best measured positions are commonly available in the optical and radio bands which guarantees the significant robustness of the resulting bound on the possible violation of WEP delivered by bias-corrected nonparametric statistical method. The forthcoming update on the data releases of \textit{Gaia} and ICRF will further improve the capability of astrometry in the test of fundamental physics and search for low-frequency gravitational wave \cite{Moore2017PRL119.261102} and local substructure of dark matter \cite{Herzog-Arbeitman2018PRL120.041102}.

\textit{Acknowledgement.} -- This work is funded by the National Natural Science Foundation of China (Grant No. 11573015). This work has made use of data from the European Space Agency (ESA) mission {\it Gaia} (\url{https://www.cosmos.esa.int/gaia}), processed by the {\it Gaia} Data Processing and Analysis Consortium (DPAC, \url{https://www.cosmos.esa.int/web/gaia/dpac/consortium}). Funding for the DPAC has been provided by national institutions, in particular the institutions participating in the {\it Gaia} Multilateral Agreement.

\bibliographystyle{apsrev4-1.bst}
\bibliography{Refs20180703}

\end{document}